\documentstyle[epsf]{elsart}
\begin{document}
GSI-Preprint-97-55, September 1997
\begin{frontmatter}
\title{Nuclear kaon dynamics}
\author[GSI]{Matthias Lutz}
%\pacs{14.20 Jn; 14.40 Aq; 25.80 Nv}
\address[GSI]{GSI, 64220 Darmstadt, Germany}

\begin{abstract}
An effective low energy Lagrangian density is applied to nuclear
$K^-$-dynamics. The free parameters, local s-wave couplings and
$SU(3)$-symmetry constrained range terms are adjusted to describe
elastic and inelastic $K^-$-nucleon scattering data. The
propagation and decay of the $\Lambda(1405)$-resonance and the
$\Lambda(1405)$-nucleon hole state is studied self consistently
with respect to the $K^-$-propagation in isospin symmetric nuclear
matter.
\end{abstract}
\end{frontmatter}

\section{Introduction}

In this letter we consider kaon propagation in isospin symmetric
nuclear matter. There has recently been much effort to evaluate the
in medium $K$-mass in realistic models. The chiral Lagrangian has
been applied perturbatively to the effective kaon mass
\cite{cpth,Min}. Brown and Rho pursue a mean field approach in
\cite{Brown}. We endeavor a microscopic description deriving the
kaon propagator in dense matter from kaon nucleon scattering in
free space \cite{Yabu,Koch,Waas1,Waas2,Pethick}. Consider the
change of the $K^+$-mass. As was emphasized in \cite{Lutz} it is
given by the low density theorem in terms of the empirical
$K^+$-nucleon scattering lengths $a_{K^+N}^{(0)}\simeq 0.02 $ fm
and $a_{K^+N}^{(1)}\simeq -0.32 $ fm \cite{Martin}
\begin{eqnarray}
\Delta \,m_{K}^2 = -\pi \left( 1+\frac{m_K}{m_N} \right)
\left( a^{(I=0)}_{KN}+3\,a^{(I=1)}_{KN} \right)
\,\rho +{\cal O}\left( k_F^4 \right)
\label{lowdensity}
\end{eqnarray}
where $\rho = 2\,k_F^3/(3\,\pi^2 )$. Model calculations
\cite{Min,Waas2} typically find only small corrections to
(\ref{lowdensity}). In fact the next to leading term of order
$k_F^4 $ can be evaluated exclusively in terms of the $K^+
N$-scattering lengths. We obtain the model independent result
\begin{eqnarray}
\Delta \,m_{K}^2 =(1)+
\alpha \left( \left(a_{K^+N}^{(I=0)}\right)^2 +3\,\left(a_{K^+N}^{(I=1)}\right)^2 \right)
 k_F^4 +{\cal O}\left( k_F^5 \right)
\label{correction}
\end{eqnarray}
where
\begin{eqnarray}
\alpha &&=\frac{1-x^2+x^2\,\log
\left(x^2\right)}{\pi^2\,(1-x)^2}
\simeq 0.166
\label{}
\end{eqnarray}
and $x =m_K/m_N $. The correction term is indeed small. At nuclear
saturation density with $k_F \simeq 265 $ MeV it increases the
repulsive $K^+$-mass shift from $28$ MeV to $ 35 $ MeV by about $20
\%$. Thus the density expansion is useful in the $K^+$-channel. Any 
microscopic model consistent with low energy $K^+$-nucleon 
scattering data is bound to give similar results for the 
$K^+$-propagation in nuclear matter at densities $\rho 
\simeq \rho_0 $ sufficiently small to maintain the density expansion rapidly
convergent. Consequently the role of chiral symmetry is restricted
to the qualitative prediction of the $K^+N$-scattering lengths by
the Weinberg-Tomozawa term.

We continue with the $K^-$-mass. Again as a first step one may
apply the low density theorem. The empirical scattering lengths
$a_{K^-N}^{(0)}\simeq (-1.70+i\,0.68 )$ fm and
$a_{K^-N}^{(1)}\simeq (0.37 +i\,0.60)$ fm \cite{Martin,Iwasaki}
imply according to (\ref{lowdensity}) a repulsive mass shift of $
23$ MeV with a width of $\Gamma_{K^{-}}
\simeq 147 $ MeV at saturation density. The correction term
(\ref{correction}) results in a total repulsive mass shift of $55 $
MeV and a width of $\Gamma_{K^-}
\simeq 195 $ MeV. At nuclear saturation the density expansion
for the $K^{-}$-mode is poorly convergent if at all. Furthermore,
the leading terms appear to contradict kaonic atom data \cite{Gal}
which suggest sizable attraction at small density. Finally the
empirical $K^-N$ scattering lengths are in striking disagreement
with the Weinberg-Tomozawa term, the leading order chiral
prediction.

The solution to this puzzle lies in the presence of the $\Lambda
(1405) $ resonance in the $K^-$-proton channel
\cite{Koch,Waas1,Kaiser}. The $\Lambda(1405)$-resonance can be
described together with elastic and inelastic $K^-$-proton
scattering data  in terms of a coupled channel Lippman-Schwinger
equation with the potential matrix evaluated perturbatively from
the chiral Lagrangian \cite{Kaiser}. We also note that a
satisfactory description of kaon nucleon scattering data can be
achieved by the coupled $K$-matrix approach of Martin
\cite{Martin}.

Contrary to the $K^+$-propagation in nuclear matter different
microscopic models consistent with low energy kaon nucleon
scattering data may predict different results for the
$K^-$-propagation in nuclear matter simply because an attractive
in-medium $K^-$-self energy probes the $K^-$- nucleon scattering
amplitude below the kaon nucleon threshold. Below the physical
threshold the amplitude is subject to uncertainties due to the
necessary subthreshold extrapolation of scattering data. Also, as
was pointed out first by Koch \cite{Koch}, the $\Lambda (1405 )$
resonance may experience a repulsive mass shift due to Pauli
blocking which strongly affects the in-medium $K^-$-nucleon
scattering amplitude \cite{Koch,Waas1}. This offers a simple
mechanism for the transition from repulsion, implied by the
scattering lengths, at low densities, $\rho < 0.1 \,\rho_0 $, to
attraction at somewhat larger densities \cite{Waas1} as favored by
kaonic atom data. Schematically this effect can be reproduced in
terms of an elementary $\Lambda (1405 )$ field dressed by  a kaon
nucleon loop. The repulsive $\Lambda $ mass shift due to the Pauli
blocking  of the nucleon is given by:
\begin{eqnarray}
\Delta\,m_\Lambda =\frac{g_{\Lambda NK}^2}{\pi^2}\,\frac{m_N}{m_\Lambda} \,
\left(1-\frac{\mu_\Lambda}{k_F}\,\arctan \left( \frac{k_F}{\mu_\Lambda} \right)
\right)k_F
\label{}
\end{eqnarray}
with the 'small' scale
\begin{eqnarray}
\mu_\Lambda^2 &=&\frac{m_N}{m_\Lambda }\left(m_K^2-\Big(m_\Lambda-m_N\Big)^2 \right)
\simeq \left( 144\, MeV \right)^2
\label{scale}
\end{eqnarray}
and the $\Lambda(1405)  $ kaon nucleon coupling constant
$g_{\Lambda NK}$.

In this letter we extend previous work \cite{Koch,Waas2} and treat
the $K^-$-state and the $\Lambda(1405)$ states self consistently.
The $\Lambda(1405) $-resonance mass in matter is then the result of
two competing effects: the Pauli blocking increases the mass
whereas the decrease of the $K^-$ mass tends to lower the mass
since the $\Lambda (1405) $  can be considered as a $K^-$-proton
bound state. We expect this mechanism to be important for in-medium
$K^-N$-scattering simply because the characteristic scale
$\mu_\Lambda $ in (\ref{scale}) depends sensitively on small
variations of $ m_\Lambda $ and $m_K$.

\section{ Kaon nucleon scattering}

We describe $K^-$ nucleon scattering by means of an effective
Lagrangian density. Consider first the isospin zero (I=0) channel:
\begin{eqnarray}
{\cal L} &=&{\textstyle{1\over 2}}\,g_{11}^{(I=0)}\,
\left( N^\dagger  \, K\right)
\left( K^\dagger \, N \right)
+{\textstyle{1\over \sqrt{6}}}\,g_{12}^{(I=0)}\,
\left(  N^\dagger \, \,K \right)
\left( \vec \pi^\dagger \cdot\vec \Sigma  \right)
\nonumber\\
&+&{\textstyle{1\over \sqrt{6}}}\,g_{21}^{(I=0)}\,
\left(  \vec \Sigma^\dagger \cdot \vec \pi \right)
\left(K^\dagger  \, N \right)
+{\textstyle {1\over 3}}\,g_{22}^{(I=0)}\,
\left(  \vec \Sigma^\dagger \cdot \vec \pi \right)
\left(  \vec \pi^\dagger \cdot \vec  \Sigma  \right)
\label{lagrangianzero}
\end{eqnarray}
with the isospin doublet fields $K=(K_{-}^\dagger,\bar K_0^\dagger)
$ and $N=(p,n)$. Here we include  the pion and the $\Sigma( 1195 )$
as relevant degrees of freedom since they couple strongly to the
$K^{-}$ nucleon system. The nucleon and $\Sigma $ as well as the
kaon and pion fields are constructed with relativistic kinematics
but without anti-particle components. The free nucleon and kaon
propagators take the form:
\begin{eqnarray}
S_N(\omega, \vec q \,) &=&\frac{m_N}{E_N(q)}\,
\frac{1}{\omega -E_N(q)+i\,\epsilon }
\nonumber\\
S_K(\omega, \vec q \,) &=&\frac{1}{2\,E_K(q )}\,
\frac{1}{\omega -E_K(q)+i\,\epsilon } \; ,
\label{}
\end{eqnarray}
respectively, where $E_a(q)=\sqrt{ m_a^2+q^2 } $. The isospin zero
coupled channel scattering amplitude
\begin{eqnarray}
T=
\left(
\begin{array}{cc}
T_{KN\rightarrow KN} & T_{KN\rightarrow \pi \Sigma } \\ T_{\pi
\Sigma \rightarrow KN } & T_{\pi \Sigma \rightarrow \pi \Sigma }
\end{array}
\right)
\label{}
\end{eqnarray}
is given by the set of ladder diagrams resumed conveniently in
terms of the Bethe-Salpeter integral equation. Since the
interaction terms in (\ref{lagrangianzero}) are local the
Bethe-Salpeter equation reduces to the simple matrix equation
\begin{eqnarray}
T(s)=g(s)+g(s)\,J(s)\,T(s) =\left( g^{-1}(s)- J(s) \right)^{-1} \;
.
\label{ts}
\end{eqnarray}
with the loop matrix $J=$ diag $\,( J_{KN}, J_{\pi
\Sigma} ) $ and
\begin{eqnarray}
J_{KN} (\omega, \vec q\,) &=&-\int_0^{\lambda } \frac{d^3
l}{(2\pi)^3 }\,
\frac{m_N}{E_N(l)}\,S_K(\omega
-E_N(l),\vec q-\vec l\,)
\nonumber\\
 J_{\pi \Sigma } (\omega ,\vec q\,) &=&-\int_0^{\lambda }
\frac{d^3 l}{(2\pi)^3 }\,
\frac{m_\Sigma}{E_\Sigma(l)}\,S_{\pi }(\omega
-E_\Sigma(l),\vec q-\vec l\,) \, .
\label{}
\end{eqnarray}
Small range terms are included in our scheme by the replacements $
g_{11} \rightarrow g_{11}+h_{11} \left( s- (m_N+m_K )^2
\right)$, $ g_{12}\rightarrow g_{12}+h_{12} \left( s- (m_N+m_K )^2 \right)
$ and $ g_{22}\rightarrow g_{22}+h_{22} \left( s- (m_\Sigma+m_\pi
)^2
\right) $ induced by appropriate additional terms in
(\ref{lagrangianzero}). The loop functions $J_{KN}$ and $J_{\pi
\Sigma }$ are regularized by the cutoff $\lambda =0.7$ GeV. For small three
momenta $|\vec q \,|< \lambda $  the loop functions $J_{KN}(\omega,
\vec q\,)$ and $J_{\pi
\Sigma}(\omega,\vec q\,)$ depend to good accuracy exclusively on the
combination $s=\omega^2-\vec q\,^2 $ as expected from covariance.

The Lagrangian density (\ref{lagrangianzero}) follows from a chiral
Lagrangian with relativistic baryon and meson fields upon
integrating out the anti-particle field components. Therefore the
coupling matrix $g$ is constrained to some extent by chiral
symmetry \cite{Kaiser}. To leading order we derive the isospin zero
coupling strengths
\begin{eqnarray}
g^{(I=0)}&=&
\left(
\begin{array}{cc} 3\,\frac{m_K}{2\,f_\pi^2} &
\sqrt{6}\,\frac{m_\pi+m_K}{8\,f_\pi^2 } \\
\sqrt{6}\,\frac{m_\pi+m_K}{8\,f_\pi^2 }&
2\,\frac{m_\pi}{f_\pi^2}
\end{array}
\right)
\label{wttermzero}
\end{eqnarray}
by matching tree level threshold amplitudes. The chiral matching of
correction terms and the range parameters $h_{ij} $ is less obvious
and not pursued here. In fact a consistent chiral matching requires
the $K^-$-nucleon potential to be evaluated minimally at chiral
order $Q^3$. Only at this order the required counter terms for the
loop functions  are introduced. In this work the coupling strengths
$g_{ij}$ and the range parameters $h_{ij} $ are directly adjusted
to reproduce empirical scattering data described in terms of the
coupled channel scattering amplitude (\ref{ts}). The  set of
parameters $ g_{11}\,
\lambda
=46.86 $, $ g_{12}\, \lambda = 11.67 $, $ g_{22}\, \lambda = 16.08 $,
$ h_{11}\,\lambda^3 = 0.79 $, $ h_{12}\, \lambda^3 = 8.57$ and
$h_{22}\,\lambda^3 =4.94 $ results from a least square fit to the
amplitudes of \cite{Kaiser}. We obtain a good description of all
coupled channel amplitudes with the isospin zero scattering length
$a^{(I=0)}_{K^-N}\simeq (-1.76 + i\,0.60) $ fm. Our parameters
confirm the result of \cite{Kaiser} that  the Weinberg-Tomozawa
term (\ref{wttermzero}) predicts the interaction strength in the
various channels rather accurately. Fig. 1 shows that the isospin
zero scattering amplitude $f^{(I=0)}_{KN}(\omega)=
m_N\,T^{(I=0)}_{KN}(\omega,\vec q=0 )/(4\,\pi \,\omega)$ is clearly
dominated by the $\Lambda(1405) $ resonance.

\begin{figure}[h]
\epsfysize=10cm
\begin{center}
\mbox{\epsfbox{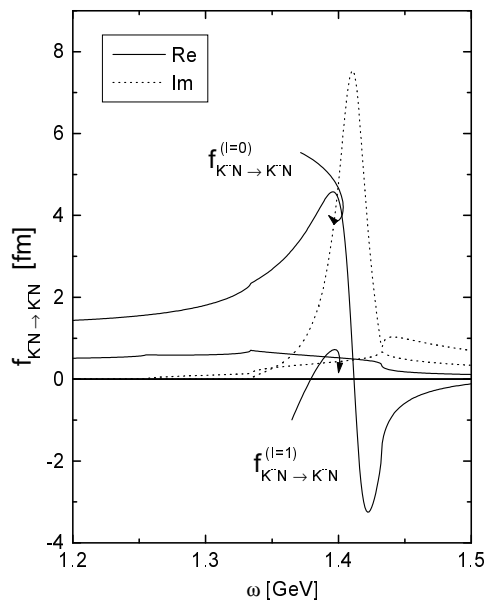}}
\end{center}
\caption{$K^-$-nucleon scattering amplitude.}
\label{fig1}
\end{figure}

Let us now turn to the $I=1$ channel. Here the $K^-$ nucleon system
couples strongly also to the $\pi\, \Lambda(1115)$-channel. The
appropriate effective Lagrangian density is:
\begin{eqnarray}
{\cal L} &=&{\textstyle{1\over 2}}\,g_{11}^{(I=1)}\,
\left( N^\dagger  \,\vec \tau \, K\right)
\left( K^\dagger \,\vec \tau\, N \right)
-{\textstyle{1\over 2}}\,g_{22}^{(I=1)}\,
\left(  \vec \Sigma^\dagger \times \vec \pi \right)
\left(  \vec \pi^\dagger \times \vec \Sigma  \right)
\nonumber\\
&+& g_{33}^{(I=1)}\,
\left( \Lambda^\dagger \vec \pi \right)
\left( \vec \pi^\dagger \, \Lambda \right)
-{\textstyle{i\over 2}}\,g_{12}^{(I=1)}\,
\Big[ \left( N^\dagger \,\vec \tau \, K \right)
\left( \vec \pi^\dagger \times \vec \Sigma \right)-h.c. \Big]
\nonumber\\
&+&{\textstyle{1\over \sqrt{2}}}\,g_{13}^{(I=1)}\,
\Big[\left( N^\dagger  \,\vec \tau \, K\right)
\left( \vec \pi^\dagger \, \Lambda \right)+h.c. \Big]
\nonumber\\
&-&{\textstyle{i\over \sqrt{2}}}\,g_{23}^{(I=1)}\,
\Big[\left( \vec \Sigma^\dagger  \times \vec \pi   \right)
\left( \vec \pi^\dagger \, \Lambda \right) -h.c. \Big]
\label{lagrangianone}
\end{eqnarray}

We construct the $I=1$ coupled channel scattering amplitude in full
analogy to the isospin zero case with $J=$ diag $(J_{KN},J_{\pi
\Sigma },J_{\pi \Lambda } )$ (see eq. (\ref{ts}) ). The range terms
are included by the replacements $ g_{11}
\rightarrow g_{11}+h_{11} \left( s- (m_N+m_K )^2
\right)$, $ g_{12}\rightarrow g_{12}+h_{12} \left( s- (m_N+m_K )^2 \right)
$, $ g_{13}\rightarrow g_{13}+h_{13} \left( s- (m_N+m_K )^2 \right)
$, $ g_{22}\rightarrow g_{22}+h_{22} \left( s- (m_\Sigma+m_\pi )^2
\right) $ and
$ g_{33}\rightarrow g_{33}+h_{33} \left( s- (m_\Lambda+m_\pi )^2
\right) $. The coupling strengths as obtained by chiral matching
of tree level threshold amplitudes are:
\begin{eqnarray}
g^{(I=1)} =
\left(
\begin{array}{ccc}
\frac{m_K}{2\,f_\pi^2} &
\frac{m_\pi+m_K}{4\,f_\pi^2 } & \sqrt{6}\,\frac{m_\pi+m_K}{8\,f_\pi^2 }
\\
\frac{m_\pi+m_K}{4\,f_\pi^2 }& \frac{m_\pi}{f_\pi^2} &0\\
\sqrt{6}\,\frac{m_\pi+m_K}{8\,f_\pi^2 } &0&0
\end{array}
\right) \; .
\label{wttermone}
\end{eqnarray}
We point out that the range terms  can be expressed in terms of the
isospin zero range parameters and one free parameter $h_F$
\begin{eqnarray}
h_{11}^{(I=1)}&=&{\textstyle{1\over2}}\,\left(h_{11}^{(I=0)}-\sqrt{6}\,h_{12}^{(I=0)}+h_{22}^{(I=0)}-6\,h_F\right)
\nonumber\\
h_{12}^{(I=1)}&=&{\textstyle{1\over6}}\,\left(3\,h_{11}^{(I=0)}+\sqrt{6}\,h_{12}^{(I=0)}-3\,h_{22}^{(I=0)}-6\,h_F\right)
\nonumber\\
h_{13}^{(I=1)}&=&{\textstyle{1\over12\sqrt{6}}}\,\left(6\,h_{11}^{(I=0)}+2\,\sqrt{6}\,h_{12}^{(I=0)}-6\,h_{22}^{(I=0)}+36\,h_F\right)
\nonumber\\
h_{22}^{(I=1)}&=&{\textstyle{1\over12}}\,\left(15\,h_{11}^{(I=0)}-5\,\sqrt{6}\,h_{12}^{(I=0)}-3\,h_{22}^{(I=0)}-30\,h_F\right)
\nonumber\\
h_{23}^{(I=1)}&=&0
\nonumber\\
h_{33}^{(I=1)}&=&{\textstyle{1\over9}}\,\left(3\,h_{11}^{(I=0)}-5\,\sqrt{6}\,h_{12}^{(I=0)}+6\,h_{22}^{(I=0)}-18\,h_F\right)
\label{su3ranges}
\end{eqnarray}
if constrained by $SU(3)$-symmetry. We obtain a good description of
the isospin one amplitudes $K^{-}N \rightarrow K^{-}N, \pi
\Sigma ,\pi \Lambda $ of \cite{Kaiser} with $g_{23}
=g_{33}=0$ as suggested by the leading order result (\ref{wttermone}).
The $K^-$-nucleon scattering amplitude $f^{(I=1)}_{K^-N}(\omega )=
m_N \,T^{(I=1)}_{K^-N}(\omega,\vec q=0 )/(4\,\pi \,\omega )$ as
shown in Fig. 1 follows with the set of parameters $g_{11}
\,\lambda =12.75 $, $g_{12}\,\lambda =13.56 $, $g_{13}\,\lambda
=15.07 $ , $g_{22}\,\lambda
= 16.04 $ and $h_F\,\lambda^3 =-1.92 $. The scattering length comes at
$a_{K^-N}^{(I=1)} \simeq (0.35 + i\, 0.69 ) $ fm. We observe that
the Weinberg-Tomozawa term (\ref{wttermone}) predicts the
interaction strengths in the $I=1$ channel less accurately than in
the $I=0$ channel. Furthermore, the imposed $SU(3)$-symmetry for
the range parameters (\ref{su3ranges}) is found to be essential for
our subthreshold extrapolation of the $K^-N$-scattering amplitude
in the $I=1$ channel.

%\newpage

\section{Kaon self energy in nuclear matter}

The kaon self energy $\Pi(\omega ,\vec q\, ) $ is evaluated in the
nucleon gas approximation. Its imaginary part follows from the
imaginary part of the in medium kaon nucleon scattering amplitude
$\bar T_{K^{-}N} $:
\begin{eqnarray}
\Im\,\Pi_K \left(\omega , \vec q \,\right)
&= &- 4\,\int_0^{k_F}\frac{d^3l}{(2\,\pi)^3}\,
\frac{m_N}{E_N(l)}\,
\Im\,\bar T_{KN}(\omega+E_N(l),\vec l+\vec q\,)
\nonumber\\
&&\;\;\;\;\;\;\;\;\;\;\;\;\;\;\;\; \cdot \,
\Theta \left( l^2-k_F^2-\omega^2+2\,\omega \,\sqrt{m_N^2+k_F^2} \,\right)
\label{kaon}
\end{eqnarray}
with the Fermi momentum $k_F $. The $\theta$-function emerges from
the zero temperature limit of appropriate Fermi-Dirac distribution
functions (see e.g. \cite{Korpa}) and ensures that the kaon
spectral density vanishes at zero energy. The real part of the self
energy then is evaluated by means of the dispersion integral:
\begin{eqnarray}
\Pi_K \left(\omega,\vec q\,\right) &=&\int_{-\infty }^{\infty }
\frac{d\,\bar \omega}{\pi}
\frac{\Im \,\Pi_K\left(\bar \omega ,\vec q \,\right)}
     {\bar \omega -\omega-i\,\epsilon }
\label{}
\end{eqnarray}
The kaon nucleon scattering amplitude $4\,\bar T_{KN}=\bar
T_{KN}^{(I=0)}+3\,\bar T_{KN}^{(I=1)}$ is medium modified
exclusively through the kaon nucleon loop since it is given by
(\ref{ts}) with the vacuum kaon nucleon loop $J_{KN} $ replaced by
the in matter loop $\bar J_{KN} $. Selfconsistency is met once
$\bar J_{KN} $ is evaluated in terms of the kaon propagator:
\begin{eqnarray}
\bar S_K(\omega , \vec q \, )
&=& \frac{1}{2\,E_K(q) }\,\frac{1}{\omega -E_K(q)-\Pi(\omega ,
\vec q \,) /(2\,E_K(q)\,)+i\,\epsilon }
\label{}
\end{eqnarray}
with the kaon self energy $\Pi_K (\omega , \vec q \, ) $ of
(\ref{kaon}). In the nucleon gas approximation the loop function
reads:
\begin{eqnarray}
\Im\,\bar J_{KN}(\omega , \vec q \,)
&&=  -\int_{k_F}^{\infty }\frac{d^3l}{(2\,\pi)^3}\,
\frac{m_N}{E_N(l)}\,
\Im\,\bar S_K(\omega-E_N(l),\vec q-\vec l\;)\,
\nonumber\\
&&\;\;\;\;\;\;\;\;\;\;\;\;\;\;\;\;\;
\cdot \Theta \left( l^2-\omega^2-m_N^2  \right)
\label{}
\end{eqnarray}
with the real part determined by the dispersion integral:
\begin{eqnarray}
\bar J_{KN}(\omega , \vec q\,) &=&\int_{-\infty}^{\infty }
\frac{d\,\bar \omega}{\pi}
\frac{\Im \,\bar J_{KN}\left(\bar \omega ,\vec q \,\right)}
     {\bar \omega-\omega-i\,\epsilon }
\label{}
\end{eqnarray}
It is convenient to formally rewrite the set of coupled equations
as follows:
\begin{eqnarray}
\bar J_{KN} \left( \omega , \vec q\, \right)
&=& -\int_{k_F}^{\lambda }\frac{d^3l}{(2\,\pi)^3}\,
\frac{m_N}{E_N(l)}\,
\bar S_K(\omega-E_N(l),\vec q-\vec l\,)
 +\Delta \,\bar J_{KN} \left(\omega , \vec q\,\right)
\nonumber\\
\Delta \,\bar J_{KN} \left(\omega ,\vec q\right)
&=&\int^{m_N}_{-\infty }\frac{d\,\bar \omega }{\pi}\,
\int_{\sqrt{\bar \omega^2-m_N^2}}^{\infty }
\frac{d^3l}{(2\,\pi)^3}\,\frac{m_N}{E_N(l)}\,
\frac{\Im\, \bar S_K\left(\bar \omega -E_N(l),\vec q-\vec l\right)}{\bar \omega-\omega -i\,\epsilon}
\label{}
\end{eqnarray}
with the kaon nucleon loop, $\bar J_{KN}(\omega , \vec q\, ) $, now
regularized by our cutoff parameter $\lambda $ such as to reproduce
the vacuum loop function $J_{KN} (\omega , \vec q, ) $ in the zero
density limit. Similarly we write
\begin{eqnarray}
\Pi_K \left( \omega , \vec q\, \right)
&= &- 4\,\int_0^{k_F}\frac{d^3l}{(2\,\pi)^3}\,\frac{m_N}{E_N(l)}\,
 \bar T_{KN}(\omega
+E_N(l),
\vec q+\vec l\,) +\Delta\,\Pi_K(\omega ,\vec q\,)
\nonumber\\
\Delta \,\Pi_K \left(\omega , \vec q\,\right)
&=&4\,\int_{-\infty }^{\mu-m_N}
\frac{d\,\bar \omega }{\pi}\,
\int_0^{\sqrt{k_F^2-2\,\bar \omega \,\mu }}\frac{d^3l}{(2\,\pi)^3}\,
\frac{m_N}{E_N(l)}\,
\frac{\Im\, \bar T_{KN}\left(\bar \omega +E_N(l),\vec q+\vec l\,\right)}{\bar \omega -\omega -i\,\epsilon}
\nonumber\\
\label{}
\end{eqnarray}
with the nucleon chemical potential $\mu^2=m_N^2+k_F^2 $. It is
immediate that $\Im\,\Delta
\,\Pi_K \left(\omega ,\vec q \right) =0 $ for $
\omega >\sqrt{m_N^2+k_F^2}-m_N $ and
$\Im\,\Delta \, \bar J_{KN} \left( \omega , \vec q\,\right) =0 $
for $\omega >m_N $. Moreover suppose $\Im \, T_{KN} (\omega )
=0 $ for $\omega < \omega_{thres.} $. The free space amplitude
of our model suggests $\omega_{thres.} = m_\Lambda+m_\pi $. Then
$\Delta\, \Pi_K =0 $ provided that $\mu< \omega_{thres. }$ or $ k_F
< \sqrt{ (m_\Lambda+m_\pi)^2-m_N^2} \simeq 830 $ MeV holds. Similar
one finds $\Delta \, \bar J_{KN} =0 $ since $\mu<
\omega_{thres. }$ implies also $ \Im \,\Pi_K(\omega )= 0$ for $\omega < 0 $.
Hence it is justified to solve our coupled set of integral
equations with $\Delta \Pi_K =\Delta \bar J_{KN}  =0 $.
\begin{figure}[h]
\epsfysize=14cm
\begin{center}
\mbox{\epsfbox{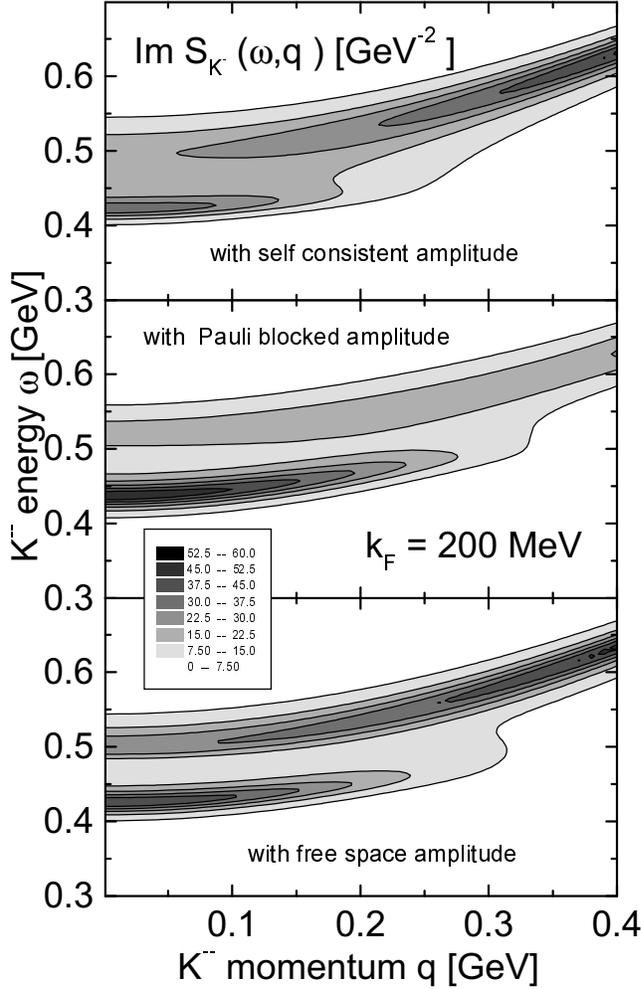}}
\end{center}
\caption{Contour plot of $K^-$-spectral density at $k_F=200 $ MeV.}
\label{fig4}
\end{figure}

We compare various degrees of approximations for the kaon self
energy. The set of equations for the kaon self energy and the
scattering amplitude is solved iteratively. We find that self
consistency is reached after few iterations: the kaon nucleon loop
function, $\bar J_{KN}$, is approximated surprisingly well if
evaluated with the kaon spectral density as obtained from the free
space scattering amplitude. Fig. 2  shows contour plots of the kaon
spectral density as evaluated with the free space kaon nucleon
scattering amplitude, with the Pauli blocked amplitude ( the
approximation scheme applied in \cite{Koch,Waas1,Waas2} ) and with
the self consistent amplitude. All three schemes predict a two mode
structure of the spectral density, however, with quantitative
differences. Both, the Pauli blocked and the self consistent
amplitude shift strength from the upper branch to the lower branch
as compared with the spectral density derived from the free space
amplitude. While the Pauli blocked amplitude causes a small
repulsive shift of the $\Lambda (1405)$-nucleon hole state the
selfconsistent amplitude predicts an attractive shift. At larger
densities self consistency affects the $K^-$-rest mass moderately.
For example at $k_F=300 $ MeV we find $\Delta \, m_{K^-}
\simeq -140 $ MeV and $\Gamma_{K^-} \simeq 35 $ MeV as compared with
$\Delta \, m_{K^-}
\simeq -141 $ MeV and $\Gamma_{K^-} \simeq 31 $ MeV from the free space
amplitude and $\Delta \, m_{K^-} \simeq -123$ MeV and $\Gamma_{K^-}
\simeq 29$ MeV from the Pauli blocked amplitude. Note here that
the quasi particle width $\Gamma_{K^- }= -\Im
\,\Pi (m_{K^-}+\Delta\, m_{K^-}, \vec q =0 )/(m_{K^-}+\Delta \, m_{K^-} )$,
given above, differs from the physical $K^-$-width by about a
factor of two due to the strong energy dependence of the kaon self
energy (see Fig. 3). We find that self consistency is most
important once the $K^-$-mode starts moving relative to the nuclear
medium. Here we expect p-wave $K^-N$-interactions
\cite{Kolomeitsev}, which are not included in this work, to modify
our results to some extent.

\begin{figure}[h]
\epsfysize=14cm
\begin{center}
\mbox{\epsfbox{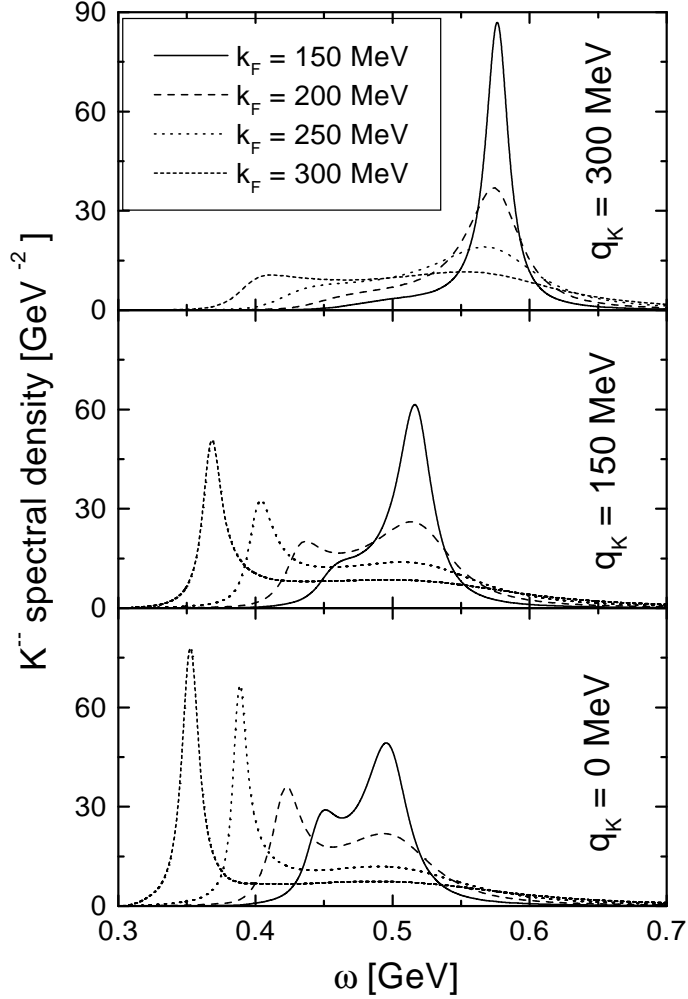}}
\end{center}
\caption{$K^-$ spectral density.}
\label{fig4}
\end{figure}
In Fig. 3 we present our final result for the kaon spectral density
as a function of the kaon energy $\omega $ for various Fermi
momenta $k_F$ and kaon momenta $\vec q $. Typically the spectral
density exhibits a two peak structure representing the $K^-$ and
the $\Lambda (1405) $-nucleon hole states. As the Fermi momentum
$k_F$ increases the energetically lower state experiences a strong
attractive shift whereas the more massive state becomes broader. On
the other hand as the kaon momentum increases both states basically
gain kinetic energy with the energetically higher peak attaining
more strength. We find that the sum rule for the kaon spectral
density
\begin{eqnarray}
\int_0^\infty  d\,\omega\,\Im \,S_K(
\omega , \vec q \,) =\frac{\pi}{2\,\sqrt{m_K^2+\vec q\,^2}}
\label{}
\end{eqnarray}
holds at the $5 \%$ accuracy level.

\begin{figure}[h]
\epsfysize=14cm
\begin{center}
\mbox{\epsfbox{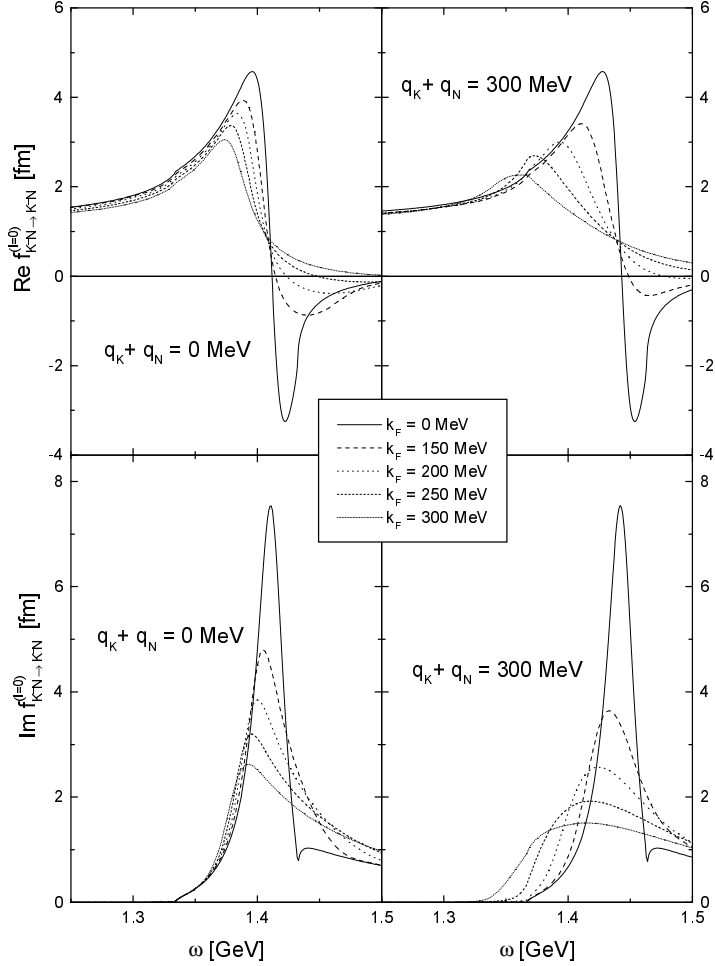}}
\end{center}
\caption{$K^-$-nucleon scattering in nuclear matter.}
\label{fig4}
\end{figure}

Fig. 4 shows the resulting isospin zero $K^-$-nucleon scattering
amplitudes with $|\vec q_K+\vec q_N| =0 $ and $|\vec q_K+\vec q_N|
=300 $ MeV. The imaginary part of the amplitude shows a clear peak
around 1.4 GeV for all Fermi momenta representing the
$\Lambda(1405) $ resonance state. As nuclear matter is compressed
the peak gets broader with little effect on the peak position. The
real part of the scattering amplitude changes its sign for $\omega
> m_N+m_K$ from repulsion to attraction as the density gets larger.

As argued in the introduction the $\Lambda(1405) $-mass shift
results from the repulsive Pauli blocking effect and the attractive
feedback effect of a decreased kaon mass. Altogether the
$\Lambda(1405)$ resonance mass is left more or less at its free
space value, however, with an increased decay width.

%\begin{figure}
%\caption{(a)
%\protect{\cite{Weinberg:nc}}.
%\mbox{$1/N_c^2$}.
%\label{elreg}
%\end{figure}

\end{document}